# Structural, optoelectronic, mechanical, thermodynamic and thermoelectric properties of kesterite $Cu_2ZnSnS_4$: A first-principles study

Iskandar Raufzoda [1,*], Amondulloi Burhonzoda[1,*], Dilshod Nematov[1,2,*], Makhmadzoir Kayumov [1,*], Saidjafar Murodzoda [1], Anushervon Ashurov [1], Sakhidod Sattorzoda [1], Makhsud Barot Islomzoda [3], Kholmirzo Kholmurodov [4,5,6], Shuwei Tang [7]

[1] S.U. Umarov Physical-Technical Institute of NAST, 734063 Dushanbe, Tajikistan
[2] Tajik Technical University named after academician M.S. Osimi. 734042 Tajikistan,
[3] Tajik National University, Dushanbe 734025, Tajikistan
[4] Frank Laboratory of Neutron Physics, Joint Institute for Nuclear Research, Dubna, Russia
[5] Department of Fundamental Nuclear Interactions, Faculty of Physics, Lomonosov Moscow State University, Moscow, Russia
[6] Department of Chemistry, New Technologies and Materials, Dubna State University, Dubna, Russia,
[7] School of Materials Science and Engineering, Liaoning Technical University, Fuxin, Liaoning 123000, China

* Corresponding authors, E-mail: raufov.i@mail.ru (IR), amondullo.burkhonzoda@mail.ru (AB), dilnem@mail.ru (DN) , kmakhmadzoir@gmail.com (MK)

## Abstract:

The structural, optoelectronic, mechanical, thermodynamic, and thermoelectric properties of kesterite $Cu_2ZnSnS_4$ were investigated using density functional theory. The mBJ and HSE06 approaches yield band gaps of 0.88 and 1.55 eV, respectively. Optical calculations reveal high absorption, with the onset in the 0.8–1.0 eV range. Hydrostatic compression from 0 to 30 kbar decreases the lattice parameters and increases the band gap. The elastic constants satisfy the Born stability criterion, and B/G > 1.75 indicates plastic behavior. Phonon calculations confirm dynamic stability. In the temperature range from 0 to 1000 K, the Helmholtz free energy decreases, the vibrational entropy increases, and the heat capacity approaches a plateau above 500 K. The thermoelectric properties were investigated in the temperature range from 100 to 1000 K. The Seebeck coefficient remains positive, peaking at approximately 310 μV/K at 250 K and decreasing to 239 μV/K at 1000 K. These results encourage the investigation of $Cu_2ZnSnS_4$ as a good candidate for thermoelectric power conversion.

## 1. Introduction

In recent decades, the search for alternative energy sources and efficient energy conversion methods has moved to the forefront of modern materials science. In this context, particular attention is being paid to materials capable of converting thermal energy directly into electrical energy (thermoelectricity) [1-3], as well as those that convert light energy into electrical energy [4-8]. Among the wide range of compounds under investigation, the kesterite phase family, specifically the quaternary semiconductor $Cu_2ZnSnS_4$ (CZTS) is considered one of the most promising candidates for future sustainable energy technologies [9-11].

Interest in CZTS stems from several factors. Traditional, highly efficient thermoelectric materials, such as bismuth telluride ($Bi_2Te_3$) and lead telluride (PbTe), exhibit impressive energy conversion performance; however, their widespread industrial application is limited by a number of significant drawbacks. First, they contain toxic components. Second, they incorporate rare and expensive elements, raising concerns regarding environmental safety and the economic viability of large-scale production [10-13]. In contrast, CZTS consists entirely of elements that are abundant in the Earth's crust, inexpensive, and environmentally benign-namely copper, zinc, tin, and sulfur. This makes the material not only economically attractive but also ideally suited for green energy concepts, where minimizing environmental impact is a mandatory requirement at every stage of a material's life cycle.

The efficiency of a thermoelectric material is traditionally evaluated using a dimensionless figure of merit, denoted as ZT. This parameter is determined by the Seebeck coefficient (which characterizes the induced thermoelectromotive force), as well as by electrical and thermal conductivity (the latter including contributions from both the crystal lattice, the phonon component and the electronic subsystem) [9, 14, 15]. High ZT values are achieved through a combination of high electrical conductivity and a high Seebeck coefficient, alongside low thermal conductivity. In this regard, CZTS also shows promising potential. Due to their complex crystal structure and the presence of heavy atoms, kesterites exhibit relatively low lattice thermal conductivity, which is an advantage for thermoelectric applications. Beyond thermoelectrics, the kesterites material $Cu_2ZnSnS_4$ has long been actively studied in the field of photovoltaics as an absorber layer material. This is due to its unique optical properties: a direct band gap that is optimal for absorbing the solar spectrum, and a high optical absorption coefficient, which allows for efficient light harvesting in thin films [9-12]. Progress in the development of CZTS-based thin-film solar cells has spurred advancements in the synthesis and characterization of this compound. However, despite extensive research into its photovoltaic properties, the thermoelectric potential of CZTS, particularly its temperature dependence, remains much less explored. There is a notable gap in our understanding of how this material behaves under illumination versus under a temperature gradient; this issue requires further analysis. From a theoretical perspective, the band structure of kesterite-type compounds creates the conditions for achieving high Seebeck coefficients, which can serve as an additional, effective tool for controlling charge carriers (for instance, through doping). At the same time, it is well known that the transport properties of semiconductors, such as electrical conductivity and thermoelectric power (Seebeck coefficient) are highly sensitive to temperature, phase transitions, and defect structure. The significant temperature dependence of the thermoelectric parameters of semiconductors necessitates their systematic analysis across various temperature ranges in order to predict the operating characteristics of devices [16]. Recent theoretical studies have shown that structural modifications can improve thermoelectric performance by regulating electron transport and reducing phonon thermal conductivity [17,18].

The search for environmentally friendly, thermodynamically stable, and accessible materials with enhanced thermoelectric properties remains a key challenge in modern materials science. In this work, we employ density functional theory (DFT) to examine the structural, electronic, optical, elastic, thermodynamic, and thermoelectric properties of $Cu_2ZnSnS_4$. Our findings provide an assessment of the compound's stability, reveal the interrelationships between its structure, electronic characteristics, and optical properties, and help determine its potential for application in thermoelectric energy conversion devices.

## 2. Computational methods

Calculations of structural, electronic, optical, and elastic properties were performed using the VASP software package [19] within the framework of density functional theory (DFT). Initially, a detailed convergence analysis regarding the cutoff energy and the k-point grid was conducted to determine the optimal parameters for subsequent computations. A cutoff energy (ENCUT) of 500 eV was selected, accounting for spin-polarization effects (ISPIN = 2). A 6×6×3 Monkhorst–Pack grid was employed for sampling the k-point space. Following the selection of appropriate computational parameters, structural optimization was carried out using the GGA-PBE, GGA-PBEsol, and SCAN approximations. Since the GGA [20] approximation is known to underestimate the band gap, more accurate functionals, namely mBJ [21], and HSE06 [22], were applied to calculate the electronic properties. The optical properties of the compound were determined based on structures optimized using the SCAN functional, with the mBJ potential applied. Elastic properties were evaluated under the same conditions using the GGA-PBE functional. Phonon and thermodynamic properties were calculated using the Phonopy package [23, 24] integrated with the VASP 6.4.3 software suite, employing the finite-displacement method for a 2×2×2 supercell. Based on the phonon density of states, the Helmholtz free energy, entropy, and heat capacity at constant volume were computed over a temperature range of 0–1000 K.

Calculations of the thermoelectric properties of $Cu_2ZnSnS_4$ were performed based on density functional theory (DFT) using the Quantum ESPRESSO software package [25]. The plane-wave basis set cutoff energy was set to 40 Rydbergs (Ry) for the wave functions and 400 Rydbergs for the charge density. Integration over the Brillouin zone was performed using a 6×6×3 k-point grid, constructed according to the Monkhorst–Pack scheme. Convergence criteria for the crystal structure optimization were set at $8.0 \times 10^{-5}$ Ry for the total energy and $1.0 \times 10^{-4}$ Ry/Bohr for the atomic forces. These computational parameters were used to optimize the geometry of the $Cu_2ZnSnS_4$ crystal structure and subsequently determine its electronic properties. To investigate the thermoelectric properties, the BoltzTraP2 software package [26] was employed, which implements a semiclassical theory of electron transport based on the Boltzmann equation. The chemical potential was set equal to the Fermi energy level, as, within the framework of the constant relaxation time approximation, both thermal and electrical conductivity depend directly on the relaxation time (τ). To obtain the final transport properties, these values can simply be multiplied by the constant relaxation time ($10^{-14}$ s, the value used in BoltzTraP2).

## 3. Results and discussion

### 3.1. Structural properties

Figure 1 shows the optimized crystal structure of tetragonal kesterite $Cu_2ZnSnS_4$. The crystal lattice is composed of Cu, Zn, Sn, and S atoms forming a three-dimensional periodic atomic network.

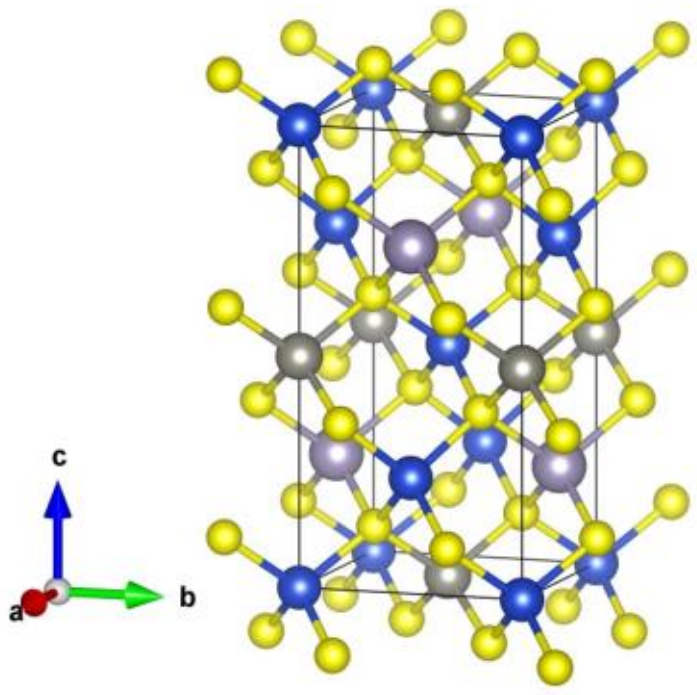

**Fig. 1.** Optimized crystal structure of kesterite $Cu_2ZnSnS_4$.

The calculated crystal lattice parameters for $Cu_2ZnSnS_4$, obtained using the GGA - PBE, GGA - PBEsol, and SCAN exchange-correlation functionals, are presented in Table 1. All three functionals predict a tetragonal crystal structure with lattice angles of $\alpha = \beta = \gamma = 90°$. All three approaches yield a tetragonal unit cell with parameters a = b and angles $\alpha = \beta = \gamma = 90°$. The GGA-PBE method yields the largest lattice parameter values, whereas PBEsol yields the smallest; the results obtained using the SCAN method fall in between. The results obtained are consistent with other literature data.

**Table 1.** Calculated lattice parameters of $Cu_2ZnSnS_4$ in Å

| **SYSTEM** | **Lattice parameters** | **This work** | | | **Other works** |
|---|---|---|---|---|---|
| | | **GGA - PBE** | **GGA - PBEsol** | **SCAN** | |
| $Cu_2ZnSnS_4$ | a, b (Å) | 5.467 | 5.374 | 5.397 | 5.42 [27], 5.48 [27] |
| | c (Å) | 10.930 | 10.752 | 10.775 | 10.89 [27], 10.85 [27] |
| | α=β=γ, ° | 90 | 90 | 90 | |
| | V (Å$^3$) | 326.65 | 310.47 | 313.89 | |

## 3.2. Electronic properties

In this section, we investigate the electronic properties of tetragonal kesterite $Cu_2ZnSnS_4$. The band gap, density of states, and band structure have been calculated. Calculating the density of states makes it possible to determine the contribution of various electronic states to the valence and conduction bands.

Based on DFT calculations using the mBJ, and HSE06 functionals, band gap values of 0.88, and 1.55 eV were obtained for $Cu_2ZnSnS_4$, respectively. The results obtained on the basis of the HSE06 functional agree very well with the experimental data given in [28], exactly by 1.5.

Figure 2 shows the calculated band structure of the $Cu_2ZnSnS_4$ compound, obtained using the mBJ approximation. The valence and conduction bands are separated by an band gap, indicating the semiconducting nature of the compound and consistent with the band gap value (0.88 eV) obtained within the mBJ approximation. The valence band maximum and the conduction band minimum are situated at the G point. Thus, the calculated spectrum indicates the presence of a direct band gap with a G→G transition. The upper group of valence bands occupies the energy range from approximately −2 to 0 eV and is characterized by a large number of closely spaced branches. Deeper valence states are observed primarily in the range from −3.7 to −3.3 eV and exhibit relatively weak dispersion.

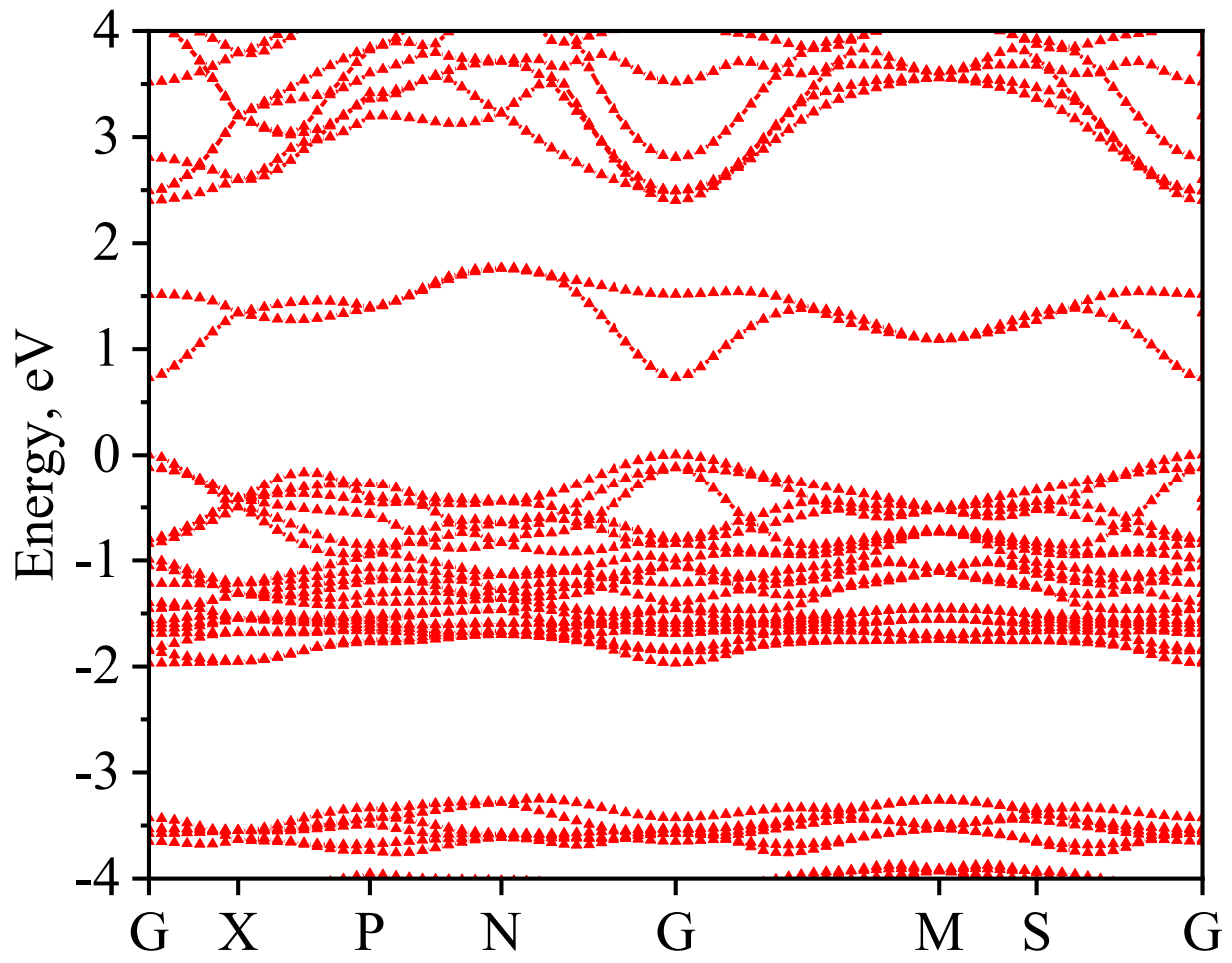

**Fig. 2.** Calculated band structure using Mbj approximation for $Cu_2ZnSnS_4$.

Figure 3 shows the total density of states for $Cu_2ZnSnS_4$ calculated using the mBJ (a) and HSE06 (b) potentials. The black and red curves correspond to spin-up and spin-down, respectively. In both approximations, the spin components of the TDOS exhibit a nearly symmetrical character. In both cases, the valence states are separated from the conduction band states by a band gap. As shown above, the band gap obtained using HSE06 is 0.67 eV larger than the value obtained using mBJ. Analysis shows that the images obtained using the mBJ and HSE06 potentials are largely similar, although the peak values of the density of states differ slightly. The maximum peak for the mBJ potential corresponds to the energy range from approximately -1.6 to -1.7 eV, while for HSE06 it falls in the range from approximately -2.6 to -2.7 eV.

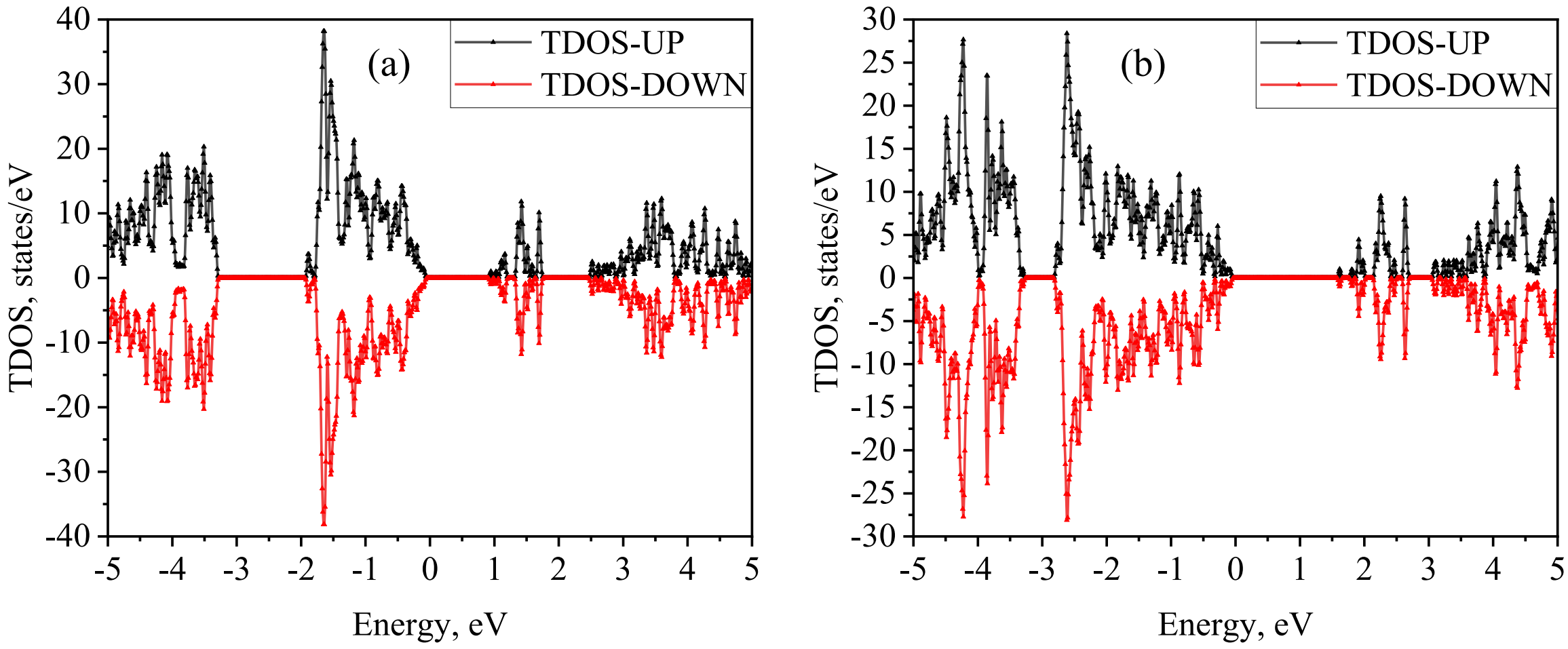

**Fig. 3.** Total density of states of $Cu_2ZnSnS_4$ calculated using (a) the mBJ potential and (b) the HSE06 hybrid functional. The Fermi level is set to 0 eV.

To investigate the contribution of each atom to the formation of the electronic states of $Cu_2ZnSnS_4$, partial densities of states were calculated using the HSE06 functional (Figure 4). The results indicate that the valence band is formed predominantly by Cu d-electrons. In the range from −1.0 eV to the Fermi level, S p-states make the dominant contribution, while the contributions from other atoms are negligible. Contributions from both types of orbitals are present in the upper part of the valence band, with sulfur p-states predominating near the band edge. Hybridization between Cu d-states and S p-states is observed.

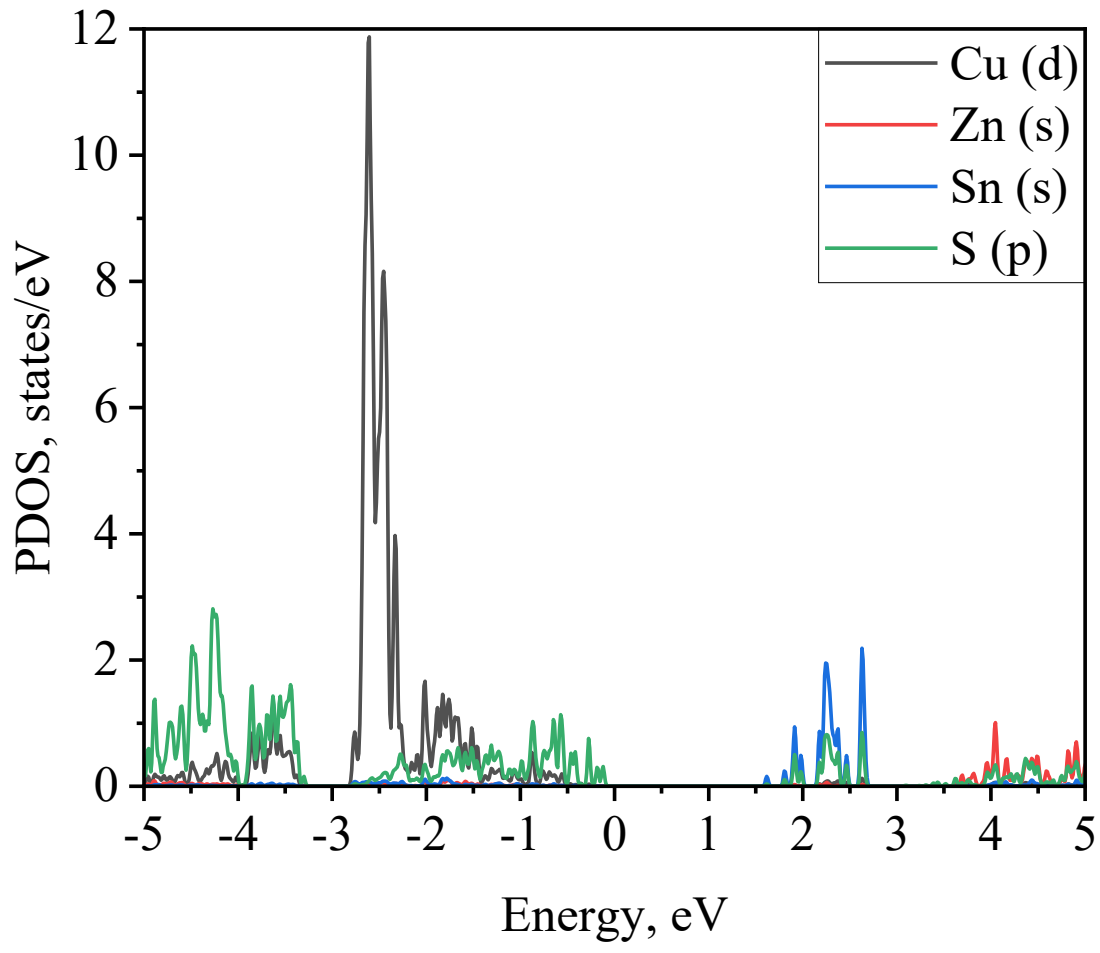

**Fig. 4.** Partial densities of states of $Cu_2ZnSnS_4$ calculated using the HSE06 hybrid functional.

Sn s-states make the primary contribution to the formation of the conduction band. The lower part of the conduction band lies in the range of approximately 1.5–1.6 eV, corresponding to the band gap value. A contribution from s-p states to the formation of the conduction band is also observed.

### 3.3. Optical Properties

Optical properties provide important insight into the electronic structure of materials and their interaction with light, enabling us to assess their suitability for use in optoelectronic devices. In this context, studying the optical properties of the kesterite material $Cu_2ZnSnS_4$ is of particular interest. These optical characteristics are described by a complex dielectric function, $\varepsilon(\omega) = \varepsilon_1(\omega) + i\,\varepsilon_2(\omega)$ [29], where $\varepsilon_1(\omega)$ and $\varepsilon_2(\omega)$ represent its real and imaginary parts, respectively. The real part characterizes the dispersive response of the material, while the imaginary part is associated with optical absorption. Based on these components, key optical properties are determined, including the absorption coefficient $\alpha(\omega)$, reflectance $R(\omega)$, refractive index $n(\omega)$, extinction coefficient $k(\omega)$, and the energy-loss function $L(\omega)$ [30–33].

$$\alpha(\omega) = \sqrt{2\omega}\left[\sqrt{\varepsilon_1^2(\omega) + \varepsilon_2^2(\omega)} - \varepsilon_1(\omega)\right]^{1/2},$$

$$R(\omega) = \frac{(n-1)^2 + k^2}{(n+1)^2 + k^2},$$

$$n(\omega) = \left[\sqrt{\varepsilon_1^2(\omega) + \varepsilon_2^2(\omega)} + \varepsilon_1(\omega)\right]^{1/2} / \sqrt{2},$$

$$L(\omega) = \frac{\varepsilon_2(\omega)}{\varepsilon_1^2(\omega) + \varepsilon_2^2(\omega)},$$

$$k(\omega) = \left[\frac{\sqrt{\varepsilon_1^2(\omega) + \varepsilon_2^2(\omega)} - \varepsilon_1}{2}\right]^{1/2},$$

Figure 5 shows the absorption coefficient α(ω), the imaginary part of the permittivity function $\varepsilon_2(\omega)$, the energy loss function $L(\omega)$, and the extinction coefficient $k(\omega)$ for $Cu_2ZnSnS_4$ as functions of energy in the range from 0 to 14 eV. Figure 5(a) shows the energy dependence of the absorption coefficient $\alpha(\omega)$ for $Cu_2ZnSnS_4$. The absorption edge begins in the energy range of 0.8–1 eV, corresponding to the band gap of this material. The first characteristic peaks appear near 2 eV. A significant increase in absorption is observed at energies above approximately 4 eV. A local maximum is located near 8 eV, and the most intense broad absorption band forms in the range of 10–11 eV, corresponding to the ultraviolet region of the spectrum. With further increase in energy, the absorption coefficient gradually decreases.

The imaginary parts of the permittivity function for $Cu_2ZnSnS_4$ are shown in Figure 5(b). It can be seen that the threshold values of $\varepsilon_2(\omega)$ are typically in the range of 0.8 eV, closely related to the band gap values. The first peak is observed near 1.8 eV, followed by a second peak at approximately 4.8 eV. The highest $\varepsilon_2(\omega)$ value is reached at approximately 7.5 eV, and then tends to 0 at higher energies.

The electron energy loss function $L(\omega)$, which quantifies the energy loss experienced by electrons during propagation through the material, is shown in Figure 5(c). Peaks in the $L(\omega)$ plots indicate that significant energy loss occurs when the incident photon energy exceeds the band gap of the material. The most noticeable increase is observed above 10 eV. At the upper limit of the range under consideration, approximately 14 eV, the value of $L(\omega)$ reaches approximately 0.52.

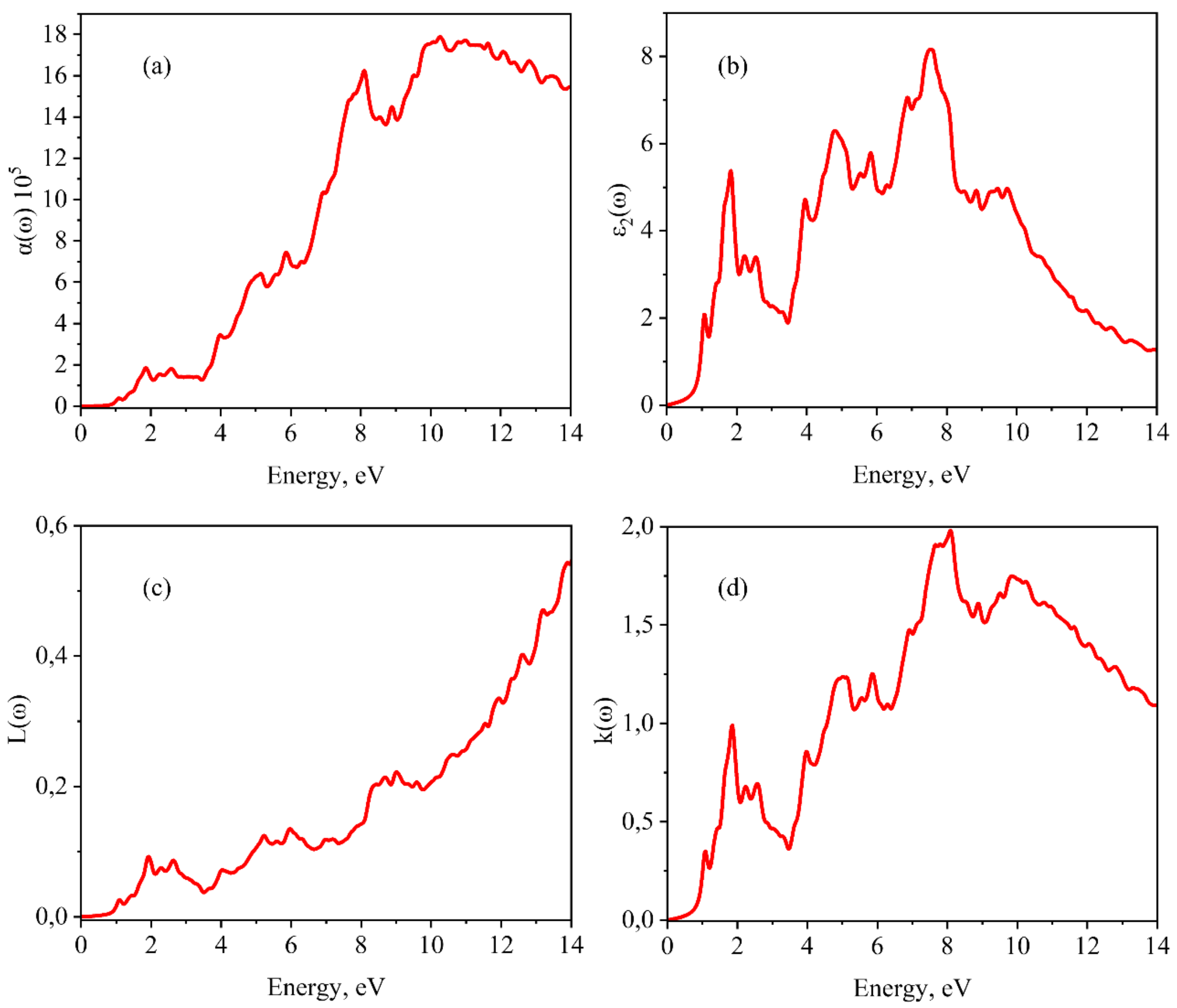

**Fig. 5.** Calculated optical properties of $Cu_2ZnSnS_4$ as functions of photon energy: (a) absorption coefficient, (b) imaginary part of the dielectric function, (c) energy loss function, and (d) extinction coefficient.

The extinction coefficient $k(\omega)$, shown in Fig. 5(d), reflects energy absorption and is similar in its properties to $\varepsilon_2(\omega)$. The first intense maximum occurs at an energy of about 2 eV, reaching a value of approximately 1.0. Following a subsequent decline, a local minimum forms near 3.5 eV, after which the value of $k(\omega)$ increases again, creating additional maxima in the 5–6 eV range. The highest value of the extinction coefficient is reached at an energy of approximately 8 eV; at higher energies, it begins to decrease.

The calculated static permittivity, represented by the real part of the permittivity function in the zero-frequency limit, $\varepsilon_1(0)$, is 7.804 for $Cu_2ZnSnS_4$. This relatively high value indicates a significant response of electron polarization to the external electric field. The corresponding static refractive index, $n(0)$, is calculated to be 2.794. The relatively high refractive index suggests a strong interaction between the electromagnetic field and the electronic structure of $Cu_2ZnSnS_4$, which is an important characteristic for its potential applications in optoelectronic and photovoltaic devices. Furthermore, the calculated zero-photon energy reflectivity, $R(0)$, is 0.224, indicating that approximately 22.4% of the incident electromagnetic radiation will be reflected from the material surface under idealized normal incidence conditions at the interface with air or vacuum. In particular, the combination of relatively high static permittivity, significant refractive index, and moderate surface reflectivity highlights the importance of optimizing light handling and minimizing reflection loss when using $Cu_2ZnSnS_4$ as an absorber material in optoelectronic devices.

### 3.4. Elastic properties

The mechanical properties of $Cu_2ZnSnS_4$ were investigated using first-principles calculations, with a particular emphasis on the pressure dependence of their elastic constants. The crystal structures were relaxed under applied pressures ranging from 0 to 30 kbar in 5 kbar increments. The calculated elastic constants provide insight into the resistance of these materials to deformation and the evolution of their elastic response under compression.

The pressure dependence of the crystal lattice parameters of tetragonal $Cu_2ZnSnS_4$ is presented in Table 2 and illustrated in Fig. 6. The values of $a = b$ and $c$ decrease monotonically with increasing pressure, reflecting the gradual compression of the crystal lattice. As the pressure increases from 0 to 30 kbar, the value of $a$ decreases from 5.467 to 5.394 Å, and the value of $c$ decreases from 10.931 to 10.788. These relative changes indicate similar compression along the crystallographic axes. Based on the data obtained, it can be said that the proportions of the unit cell are largely preserved under compression.

**Table 2.** Calculated lattice constants of $Cu_2ZnSnS_4$ under pressures ranging from 0 to 30 kbar within the framework of the GGA.

| **System** | **$Cu_2ZnSnS_4$** | |
|---|---|---|
| **Pressure (kbar)** | ***a, b*** **(Å)** | ***c*** **(Å)** |
| 0 | 5.467 | 10.931 |
| 5 | 5.453 | 10.905 |
| 10 | 5.441 | 10.880 |
| 15 | 5.428 | 10.855 |
| 20 | 5.416 | 10.832 |
| 25 | 5.405 | 10.809 |
| 30 | 5.394 | 10.788 |

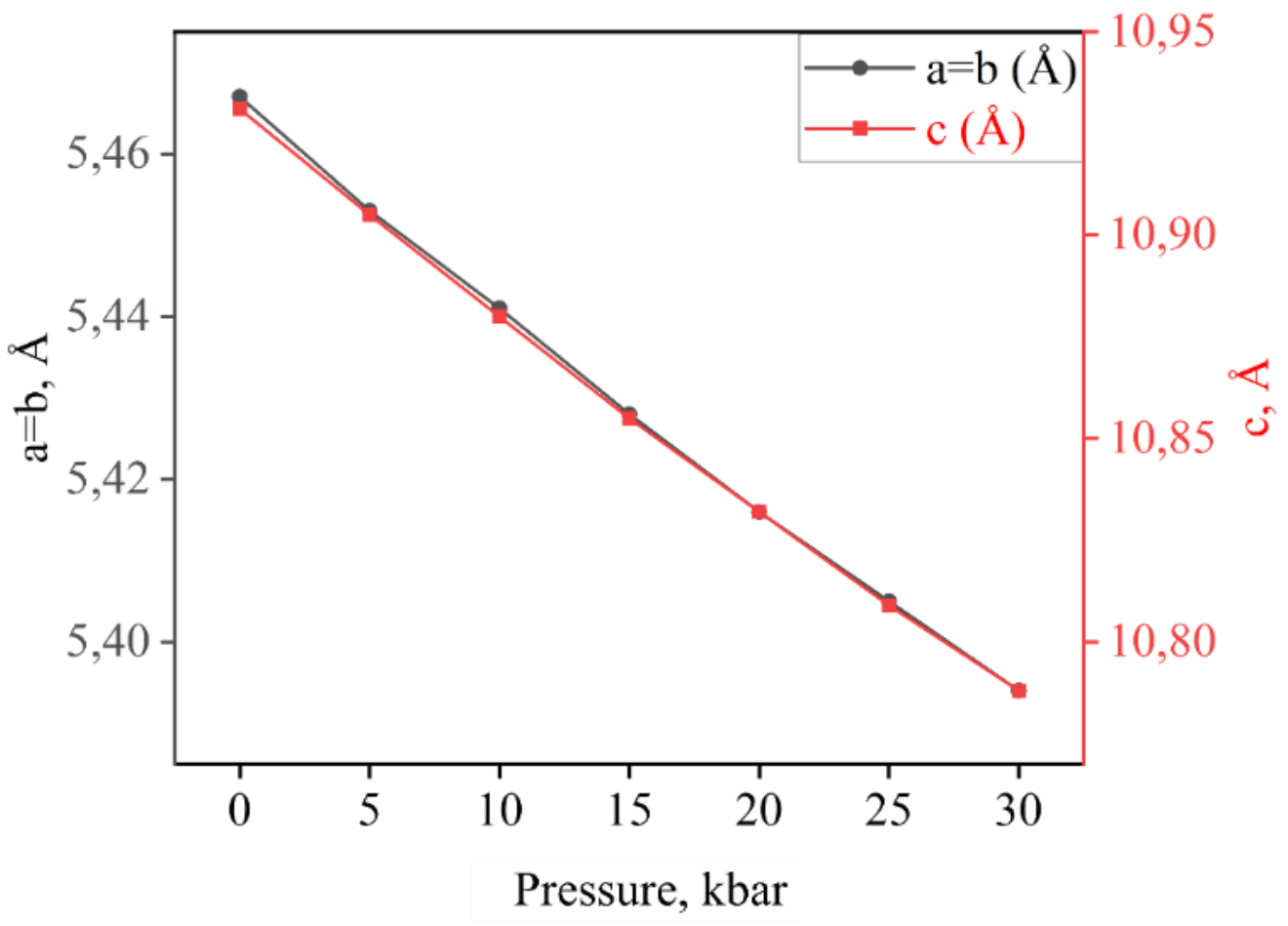

**Fig. 6.** Calculated lattice constants of $Cu_2ZnSnS_4$ as a function of pressure.

Next, the electronic properties of $Cu_2ZnSnS_4$ under pressure were calculated. The obtained results are presented in the form of the band gap value in Table 3 and plotted in Figure 7 as the dependence of the band gap on the applied external pressure. The obtained results show that the band gap of $Cu_2ZnSnS_4$ increases with increasing pressure, increasing from 0.5825 eV at 0 kbar to 0.6727 eV at 30 kbar. These results indicate a gradual expansion of the band gap under pressure, while the calculated electronic structure retains its semiconductor character throughout the studied range.

**Table 3.** Calculated band gap of $Cu_2ZnSnS_4$ under pressures ranging from 0 to 30 kbar within the framework of the GGA.

| System | Pressure (kbar) | Band Gap |
|---|---|---|
| $Cu_2ZnSnS_4$ | 0 | 0.5825 |
| | 5 | 0.5988 |
| | 10 | 0.6136 |
| | 15 | 0.6291 |
| | 20 | 0.6440 |
| | 25 | 0.6585 |
| | 30 | 0.6727 |

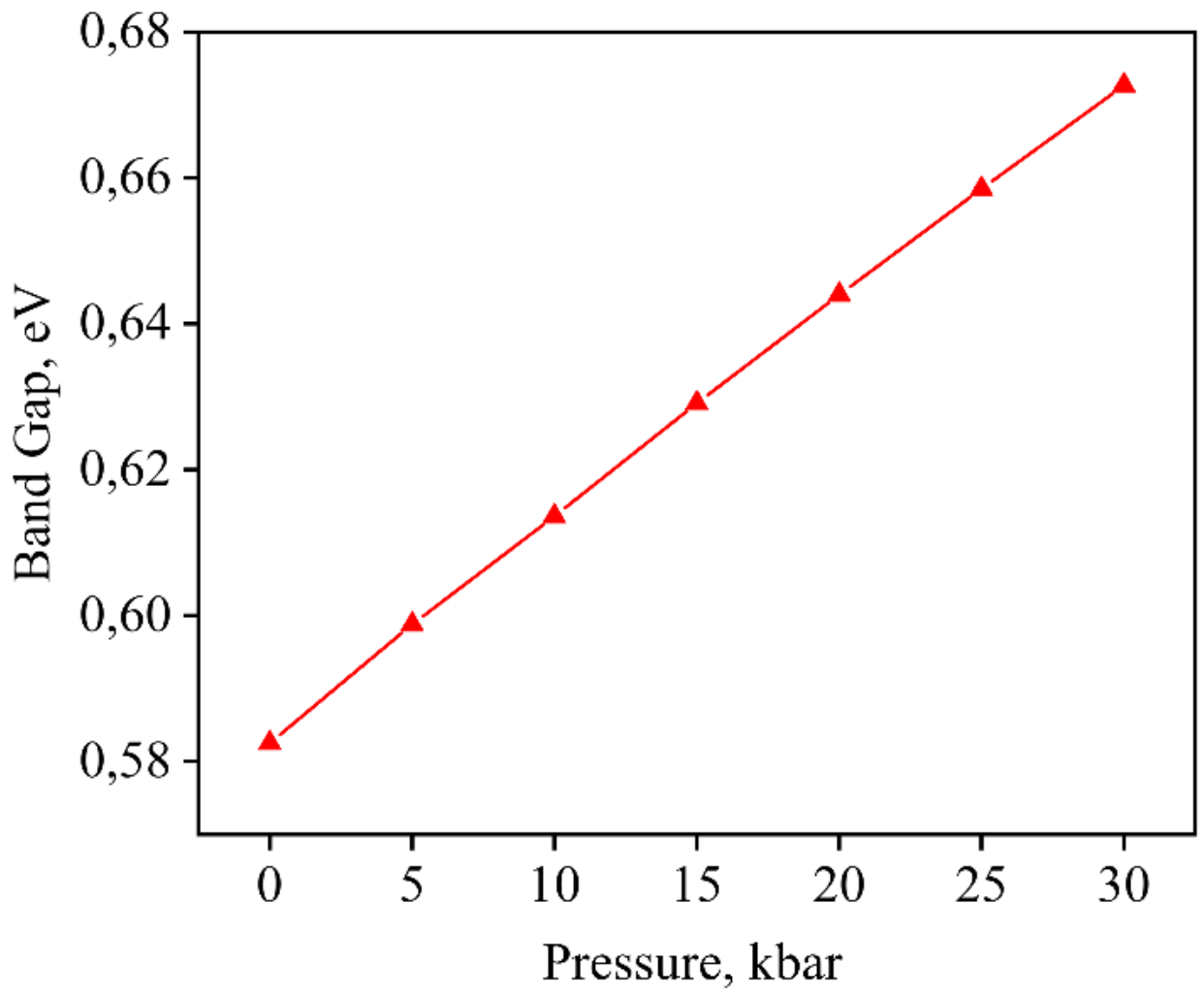

**Fig. 7.** Calculated band gap of $Cu_2ZnSnS_4$ as a function of pressure.

The mechanical properties of $Cu_2ZnSnS_4$ were investigated using first-principles calculations, with a particular emphasis on the pressure dependence of their elastic constants. The crystal structures were relaxed under applied pressures ranging from 0 to 30 kbar in 5 kbar increments. The calculated elastic constants provide insight into the resistance of these materials to deformation and the evolution of their elastic response under compression.

Elastic constants characterize the rigidity of a crystal under external deformation and allow us to understand the mechanical stability of its structure [34].

The elastic stability of tetragonal $Cu_2ZnSnS_4$ was evaluated using VASPKIT [35] based on the criteria proposed by Muchat and Coudert [36]. For tetragonal srtuctures, the analysis took into account 7 independent elastic constants: $C_{11}$, $C_{12}$, $C_{13}$, $C_{16}$, $C_{33}$, $C_{44}$, and $C_{66}$. The following stability inequalities were investigated:

$$C_{11} > |C_{12}|,$$

$$C_{33}(C_{11} + C_{12}) > 2C_{13}^2,$$

$$C_{44} > 0,$$

$$C_{66}(C_{11} - C_{12}) > 2C_{16}^2.$$

The values obtained in this study satisfy these fundamental sustainability requirements. The calculated elastic constants of tetragonal $Cu_2ZnSnS_4$, presented in Table 4 and depicted in Figure 8, respond differently to the applied pressure.

**Table 4.** Pressure dependence of the elastic constants $C_{ij}$ (all in GPa) and bulk moduli *B* (GPa).

| System | Pressure, kbar | $C_{11}$ | $C_{12}$ | $C_{13}$ | $C_{16}$ | $C_{33}$ | $C_{44}$ | $C_{66}$ | *B* |
|---|---|---|---|---|---|---|---|---|---|
| $Cu_2ZnSnS_4$ | 0 | 89.880 | 53.648 | 53.593 | 1.354 | 89.529 | 40.930 | 39.939 | 65.538 |
| | 5 | 91.729 | 55.524 | 55.414 | 1.398 | 91.005 | 40.767 | 39.977 | 67.294 |
| | 10 | 93.267 | 57.414 | 57.447 | 1.577 | 93.136 | 40.822 | 39.987 | 69.256 |
| | 15 | 95.192 | 59.697 | 59.588 | 1.470 | 95.005 | 40.840 | 39.964 | 71.396 |
| | 20 | 96.742 | 61.450 | 61.383 | 1.427 | 96.606 | 40.827 | 39.917 | 73.102 |
| | 25 | 97.741 | 62.771 | 63.250 | 1.547 | 98.566 | 40.652 | 39.812 | 74.572 |
| | 30 | 98.818 | 64.192 | 64.859 | 1.687 | 100.256 | 40.577 | 39.740 | 76.074 |

With increasing pressure from 0 to 30 kbar, the values of $C_{11}$, $C_{33}$, $C_{12}$, and $C_{13}$ increase. The values of $C_{44}$ and $C_{66}$ exhibit a relatively weak dependence on pressure with small nonmonotonic variations. At the same time, the elastic modulus *B* increases steadily from 65.538 GPa at atmospheric pressure to 76.074 GPa at 30 kbar, which is an increase of approximately 16.1% and indicates greater resistance to volumetric compression.

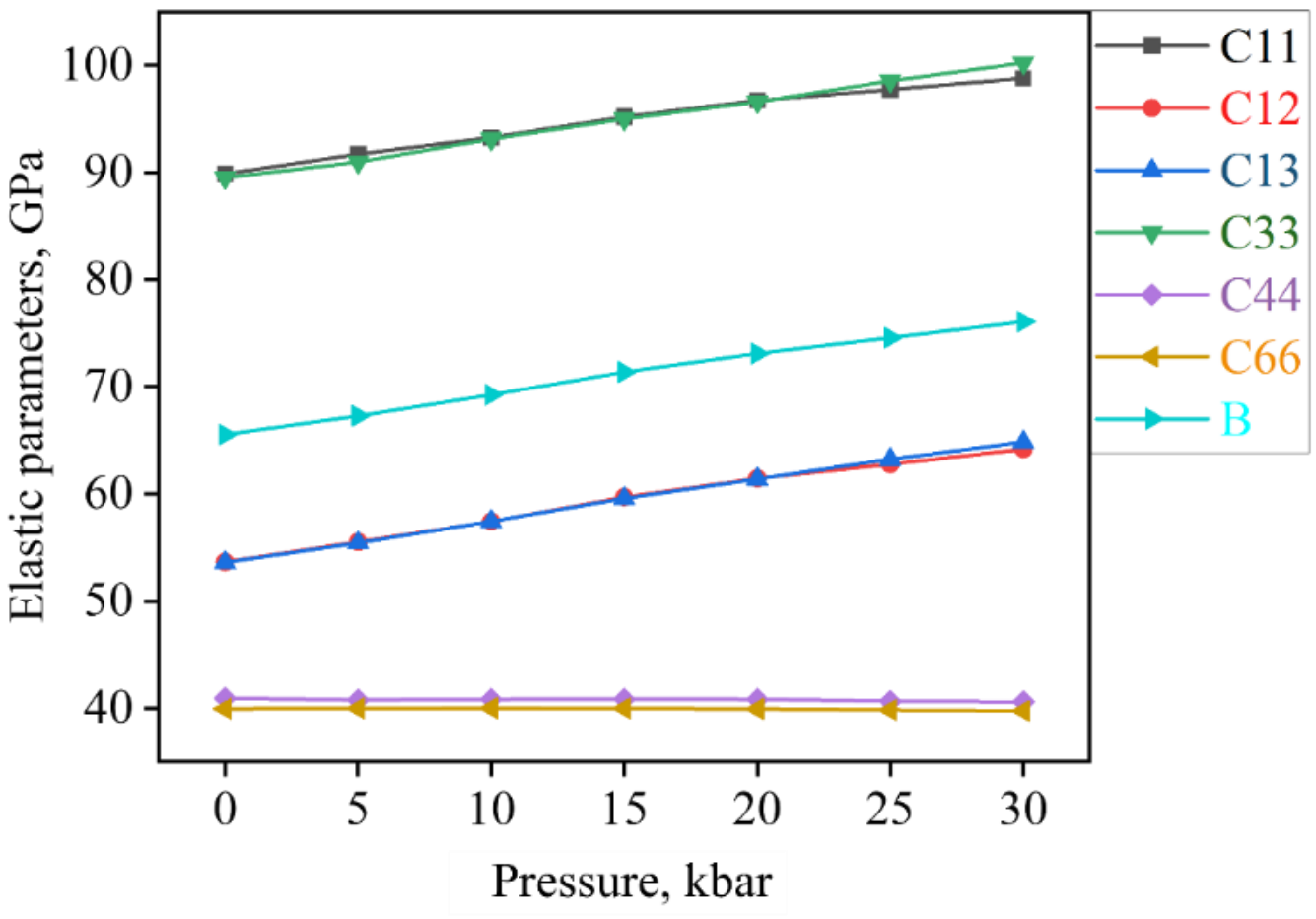

**Fig. 8.** Calculated elastic constants $C_{ij}$ and bulk moduli $B$ for $Cu_2ZnSnS_4$ as a function of pressure.

Figure 9(a) shows the dependence of the shear modulus and sound velocity (Figure 9(b)) on pressure in $Cu_2ZnSnS_4$. The shear modulus decreases with increasing pressure. This decrease becomes more pronounced at pressures above 20 kbar, indicating a gradual decrease in shear strength. The longitudinal sound velocity, $v_l$, increases slightly with pressure, while the transverse and average sound velocities, $v_t$ and $v_m$, gradually decrease. The order $v_l > v_m > v_t$ is maintained over the entire pressure range studied. Overall, the obtained results demonstrate that compression affects the bulk and shear properties of $Cu_2ZnSnS_4$ differently.

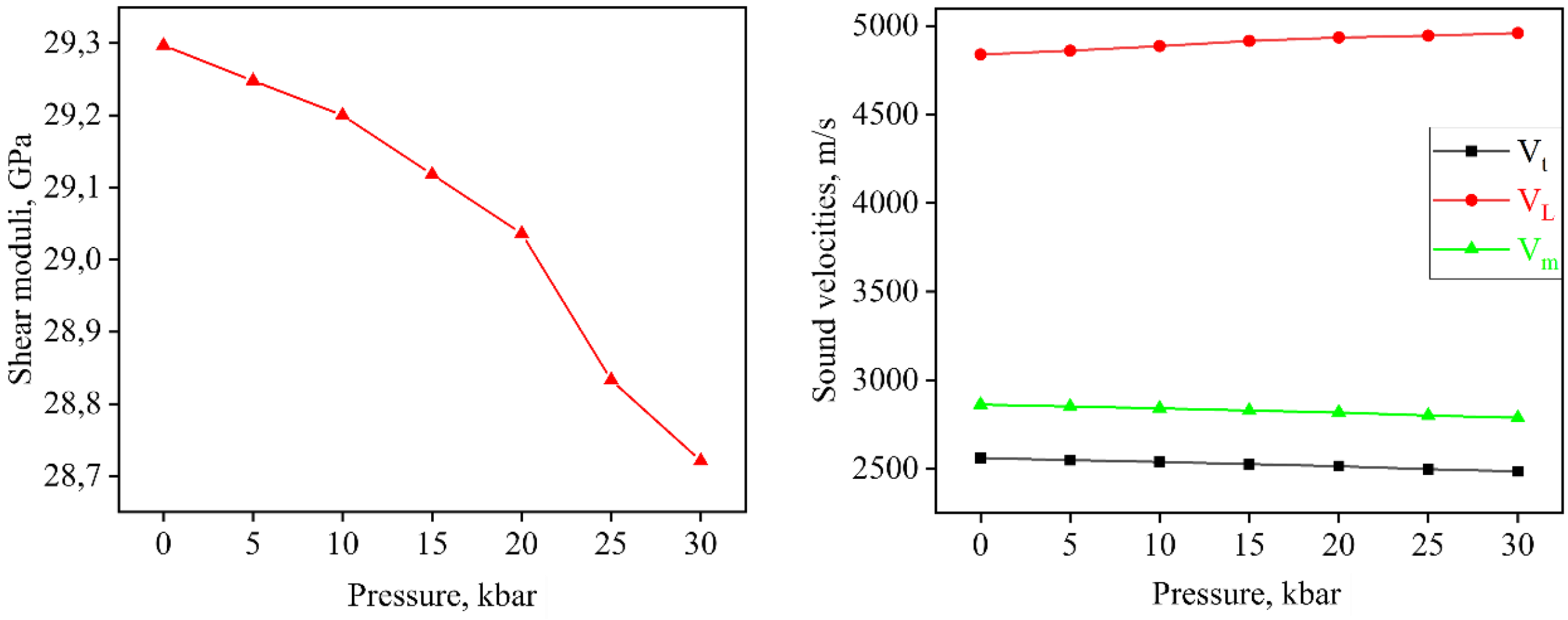

**Fig. 9.** Calculated (a) shear moduli and (b) sound velocities of $Cu_2ZnSnS_4$ as a function of pressure.

The Debye temperature is sensitive to various physical parameters, such as melting point, phonon heat capacity, and elastic constants. The calculated Debye temperature as a function of pressure is shown in Fig. 10.

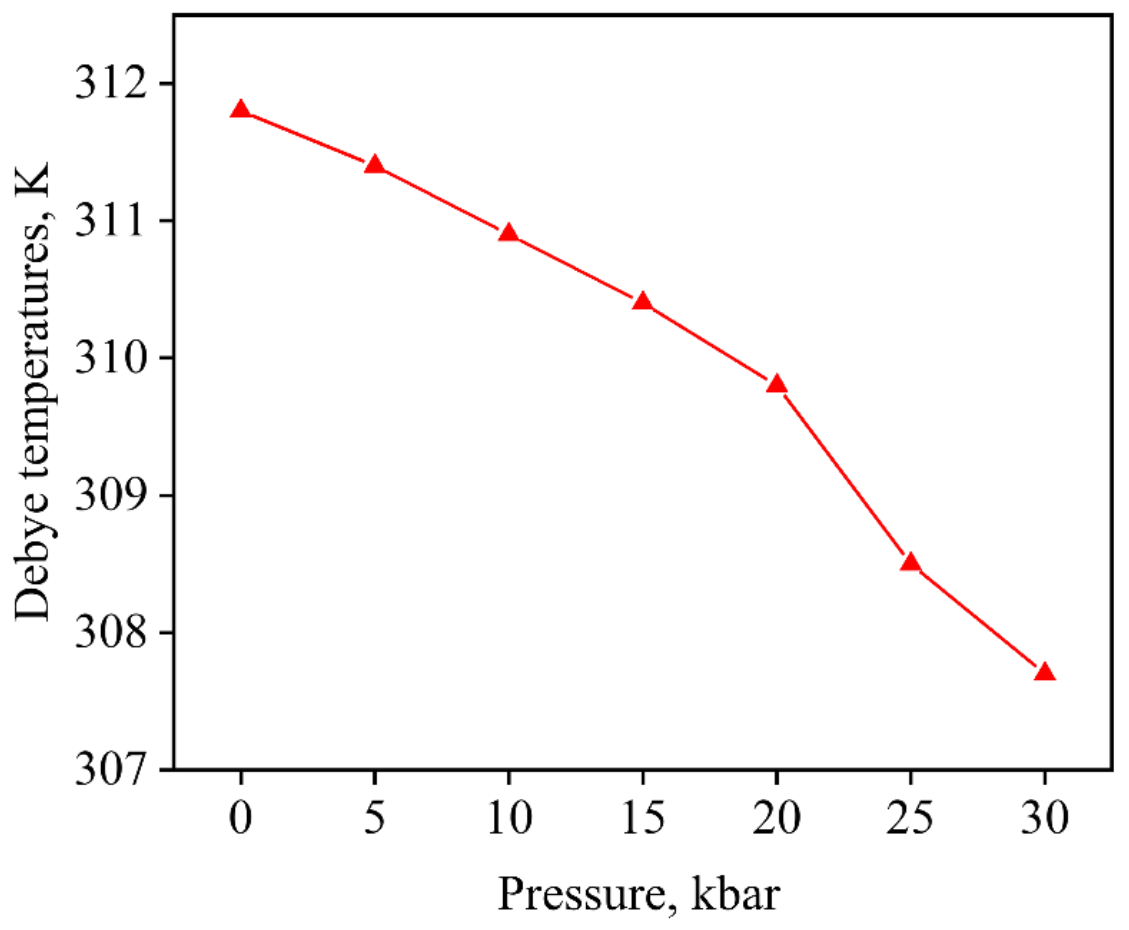

**Fig. 10.** Calculated Debye temperature of $Cu_2ZnSnS_4$ as a function of pressure.

The Debye temperature decreases with increasing pressure, from approximately 311.9 K at 0 kbar to 307.8 K at 30 kbar. This corresponds to an overall decrease of approximately 1.3%, indicating a relatively weak dependence on pressure in the studied range. A more pronounced decrease in the Debye temperature is observed in the range from 20 to 30 kbar. Table 5 presents the calculated values of Young's modulus, Poisson's ratio, and Puig's ratio for $Cu_2ZnSnS_4$ at pressures from 0 to 30 kbar in increments of 5 kbar. The obtained data demonstrate a weak dependence of Young's modulus on pressure. Young's modulus increases up to 20 kbar and then decreases slightly with increasing pressure. Overall, the change is minor and indicates that the effective stiffness under load is largely maintained over the entire range of applied pressures.

**Table 5.** Pressure dependence of the Young's Modulus, Poisson's Ratio and Pugh's Ratio.

| Mechanical properties / System | P, kbar | Young's Modulus (GPa) | Poisson's Ratio | Pugh's Ratio |
|---|---|---|---|---|
| $Cu_2ZnSnS_4$ | 0 | 76.492 | 0.305 | 2.237 |
| | 5 | 76.640 | 0.310 | 2.301 |
| | 10 | 76.805 | 0.315 | 2.372 |
| | 15 | 76.900 | 0.320 | 2.452 |
| | 20 | 76.922 | 0.325 | 2.518 |
| | 25 | 76.623 | 0.329 | 2.586 |
| | 30 | 76.532 | 0.332 | 2.649 |

Poisson's and Pugh's ratios also increase with increasing pressure. This trend indicates an increase in longitudinal stiffness with limited lateral deformation. All calculated Pugh's ratio values exceed the generally accepted empirical threshold of 1.75, allowing $Cu_2ZnSnS_4$ to be classified as a ductile material in the Pugh's classification system across the entire test range. Thus, an increase in the Pugh's coefficient indicates an increasing tendency towards plastic behavior under compression.

### 3.5. Thermodynamic properties

Next, the thermodynamic properties of $Cu_2ZnSnS_4$ were analyzed to evaluate its dynamic stability and thermal stability. Figure 11(a) shows the phonon dispersion relations for $Cu_2ZnSnS_4$. The acoustic branches approach zero frequency at point G, as expected for lattice translational vibrations, while the optical branches maintain finite frequencies. Several optical branches exhibit relatively weak dispersion, resulting in pronounced features in the phonon density of states. The absence of imaginary frequencies confirms the dynamic stability of the compounds. Figure 11(b) shows the phonon density of states for the studied compounds. The overall phonon density of states exhibits two distinct frequency regions. The low-frequency region contains pronounced peaks near 2 and 5 THz, while the high-frequency region includes several closely spaced peaks in the range from approximately 7.5 to 10.5 THz.

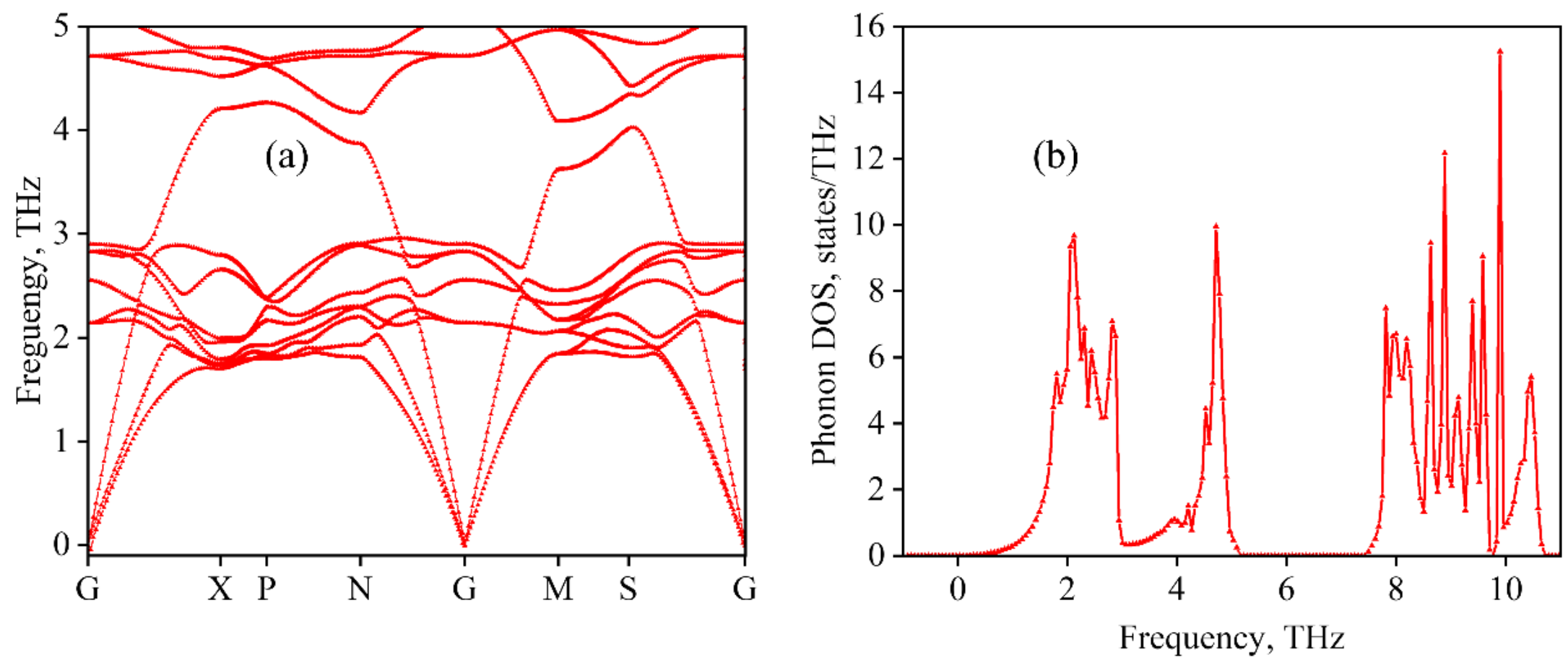

**Fig. 11.** Calculated (a) phonon dispersion relations and (b) phonon density of states for $Cu_2ZnSnS_4$.

Figure 12 shows the temperature-dependent thermodynamic properties of $Cu_2ZnSnS_4$, including the free energy, entropy, and heat capacity at constant volume, in the temperature range of 0–1000 K.

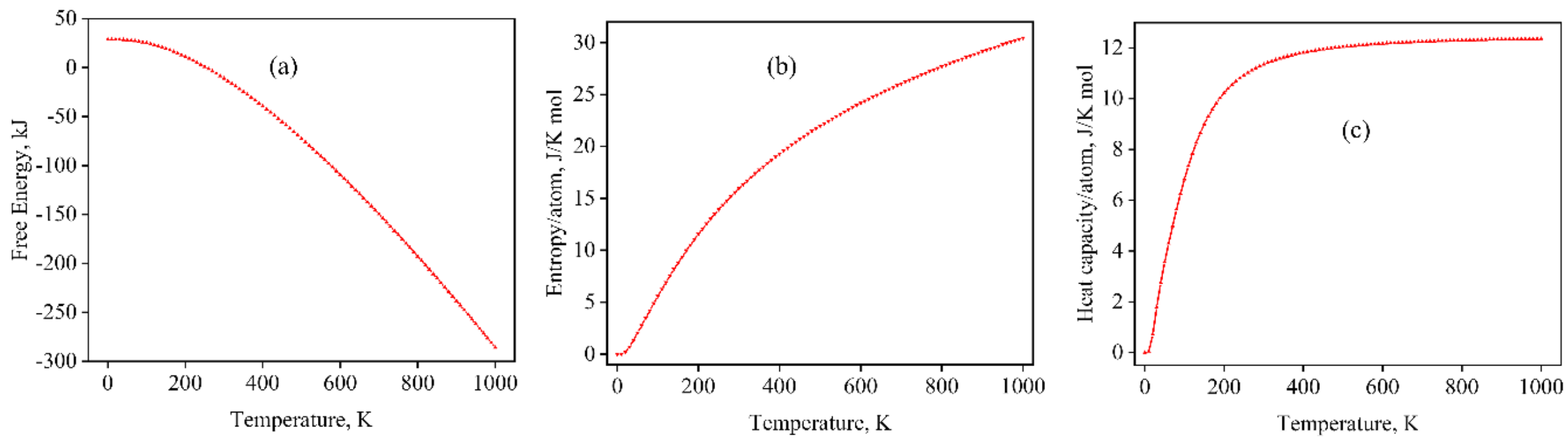

**Fig. 12.** Temperature dependence of the thermodynamic properties of $Cu_2ZnSnS_4$: (a) free energy, (b) entropy, and (c) heat capacity.

As shown in Figure 12(a), the Helmholtz free energy decreases monotonically with increasing temperature. With heating, the increasing entropy contribution gradually reduces the free energy. At the same time, the vibrational entropy increases continuously from approximately zero, as shown in Figure 12(b), reflecting the increasing thermal depletion of vibrational states.

The heat capacity, shown in Figure 12(c), approaches zero at low temperatures because thermal excitation of lattice vibrations is strongly suppressed. With heating, it increases rapidly as the number of thermally phonon modes increases. Above approximately 500 K, its temperature dependence becomes relatively weak. These trends characterize the vibrational-thermodynamic response of $Cu_2ZnSnS_4$.

### 3.6. Thermoelectrical properties

In recent years, the demand for energy has been increasing, accompanied by significant heat loss in industrial and energy-related processes. This has drawn the attention of researchers toward the development of efficient materials capable of converting thermal energy into electrical energy. In this context, thermoelectric materials are of particular importance, as they offer the potential to convert waste heat directly into electrical energy. Despite significant progress in this field, the development of materials that combine high efficiency, chemical and thermal stability, and minimal environmental impact remains a key scientific and technological challenge [37-41]. $Cu_2ZnSnS_4$ is composed of abundant, non-toxic elements and allows for the tuning of its electronic properties, making it a promising compound for use in thermoelectric devices.

The thermoelectric properties of $Cu_2ZnSnS_4$ were evaluated using the BoltzTraP software package, based on an energy band structure calculated via the GGA-PBE method. The temperature dependence of thermoelectric properties - such as the Seebeck coefficient (S), power factor (PF), charge carrier concentration, thermal conductivity ($\kappa_e$), and electrical conductivity ($\sigma$) - was investigated within the temperature range of 0 to 1000 K. The temperature dependence of the charge carrier concentration is illustrated in Figure 13. In the 100–200 K range, the charge carrier concentration remains very low, which is characteristic of semiconductors at low temperatures. As the temperature rises, a sharp increase is observed, attributed to the thermal excitation of electrons from the valence band to the conduction band. At 1000 K, the charge carrier concentration reaches approximately $109 \times 10^{-3}$ e/uc. Absolute values for the charge carrier concentration are not provided, as the calculations were performed under the constant relaxation time approximation (CRTA) and are intended for the analysis of temperature-dependent trends.

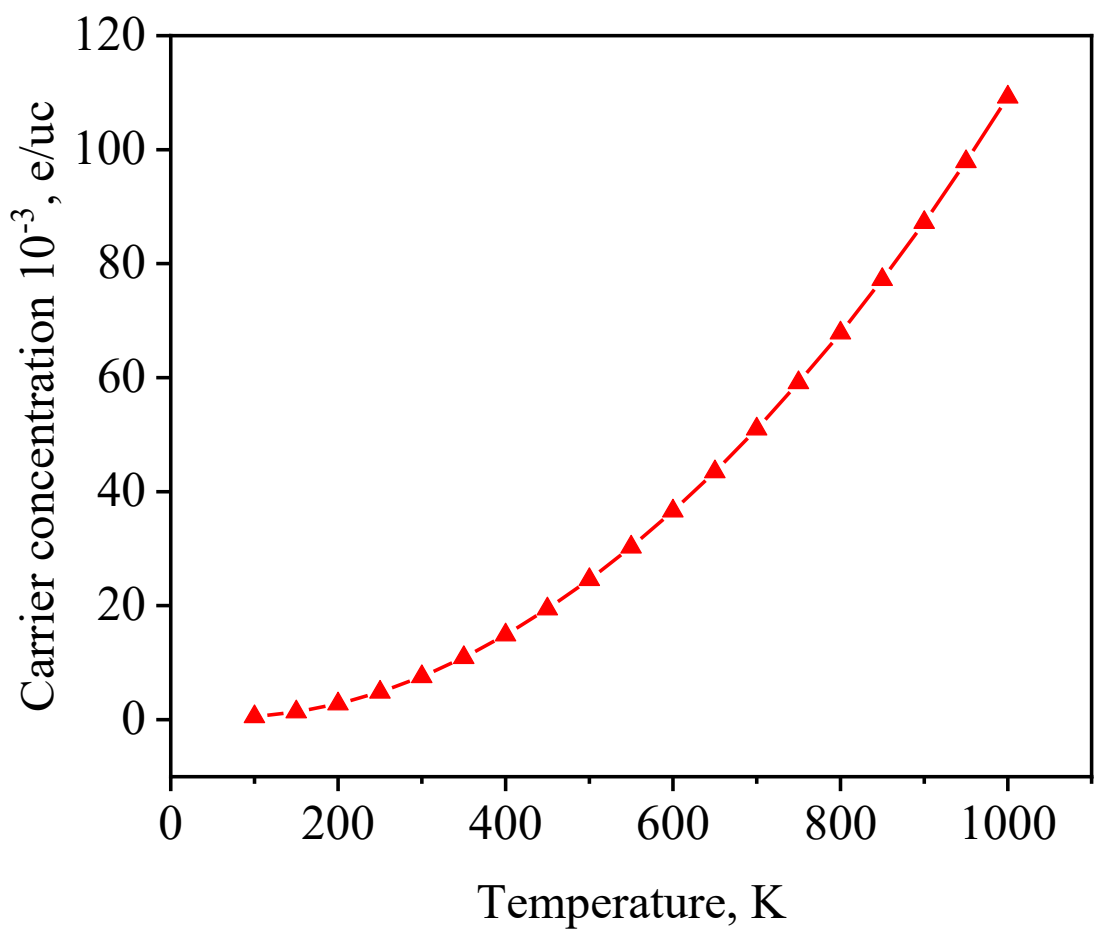

**Fig. 13.** Temperature dependence of charge carrier concentration in $Cu_2ZnSnS_4$

An increase in the charge carrier concentration has a significant effect on the Seebeck coefficient, the temperature dependence of which is shown in Figure 14. At a temperature of 100 K, the Seebeck coefficient is 264 μV/K, reaching a maximum value of approximately 310 μV/K at 250 K. With a further increase in temperature, S gradually decreases, reaching 239 μV/K at 1000 K. A positive value of S indicates the presence of p-type conduction in $Cu_2ZnSnS_4$. This is associated with an increase in the charge carrier concentration and is consistent with the relationship $S \propto 1/n$, which is characteristic of intrinsic semiconductors.

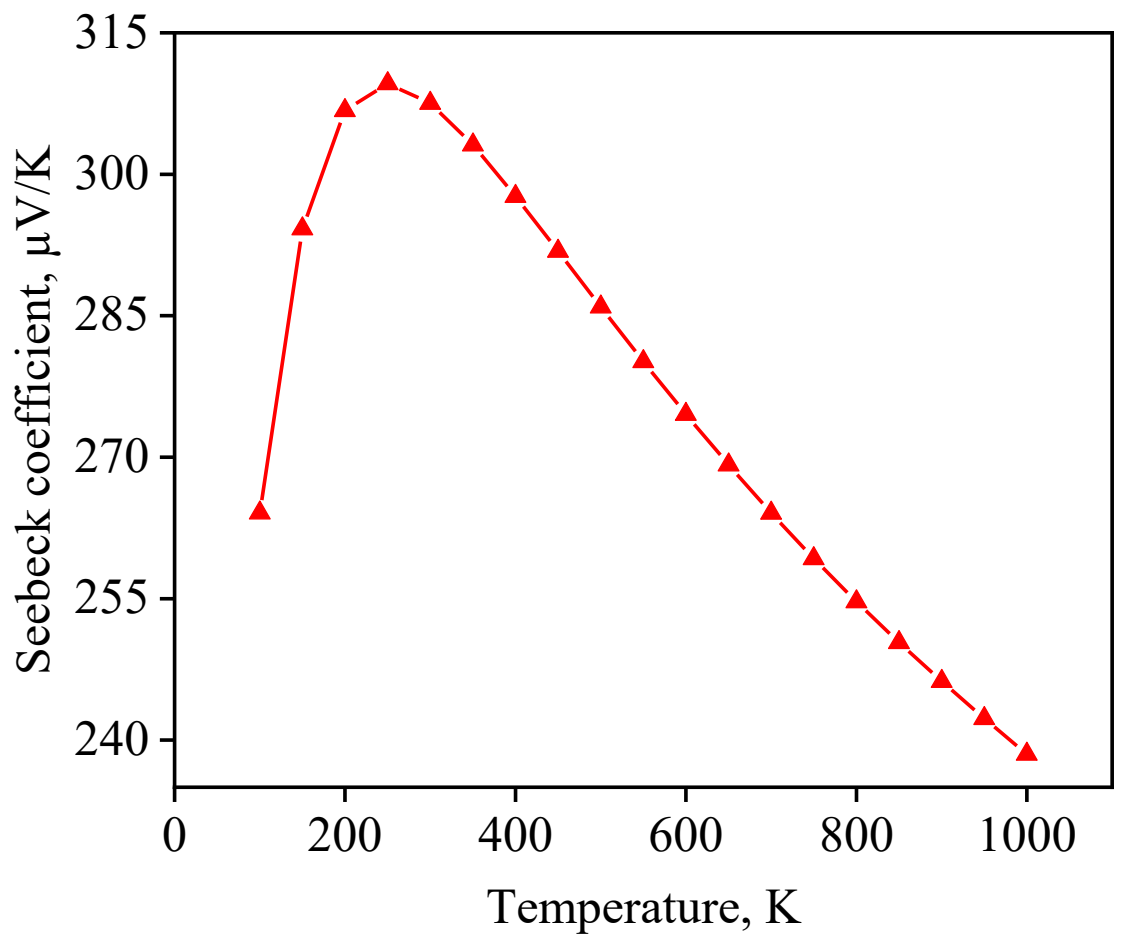

**Fig. 14.** Seebeck coefficient as a function of temperature for $Cu_2ZnSnS_4$.

Figure 15 shows the temperature dependence of the electron thermal conductivity $\kappa_e$. At low temperatures (up to ~200 K), the contribution of electrons to thermal conductivity is negligible. Beginning at approximately 250 K, the electron thermal conductivity increases and continues to increase with temperature due to the increasing charge carrier concentration. At room temperature, $\kappa_e$ reaches a value of approximately 0.082 W/mK. The increase in current flow with temperature confirms the semiconducting nature of the material.

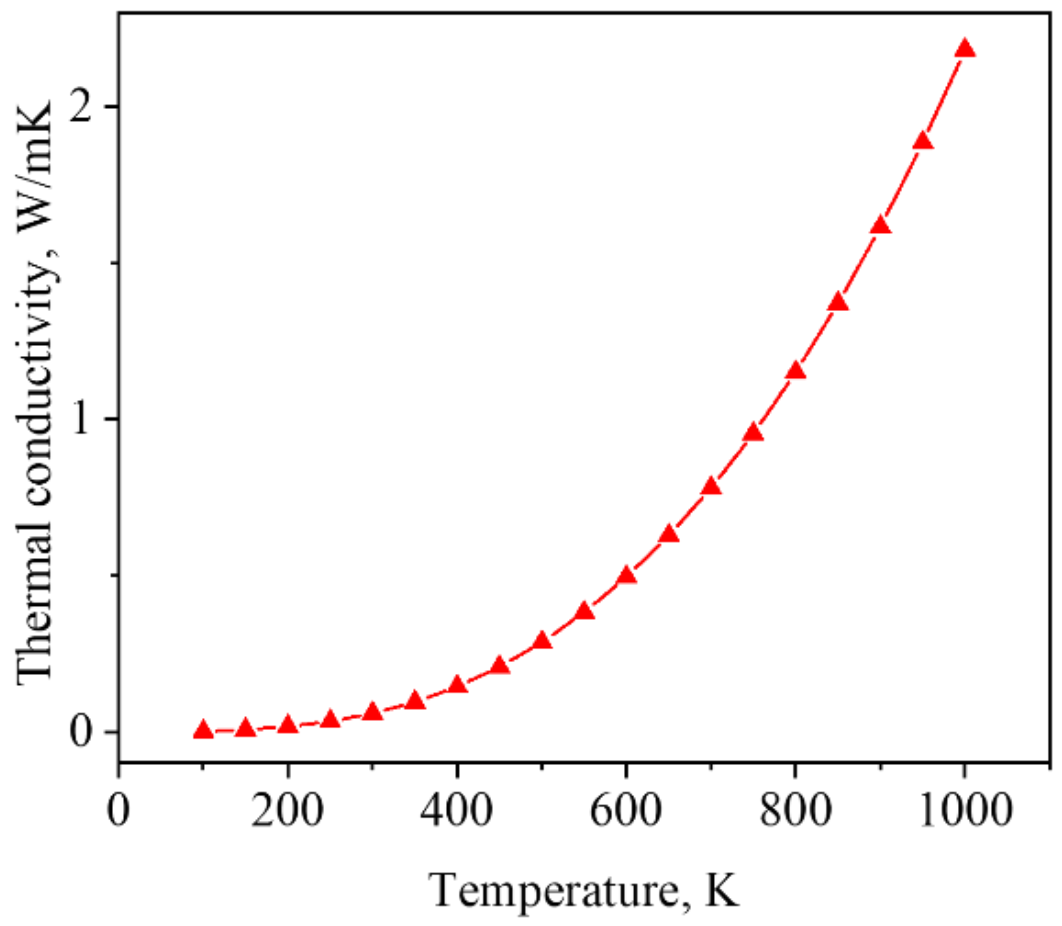

**Fig. 15.** Temperature dependence of electron thermal conductivity for $Cu_2ZnSnS_4$.

The temperature dependence of electrical conductivity σ is shown in Figure 16. At low temperatures, electrical conductivity is low due to the low charge carrier concentration. In the range of 300–1000 K, a sharp increase in σ is observed, caused by the transition of charge carriers to the conduction band.

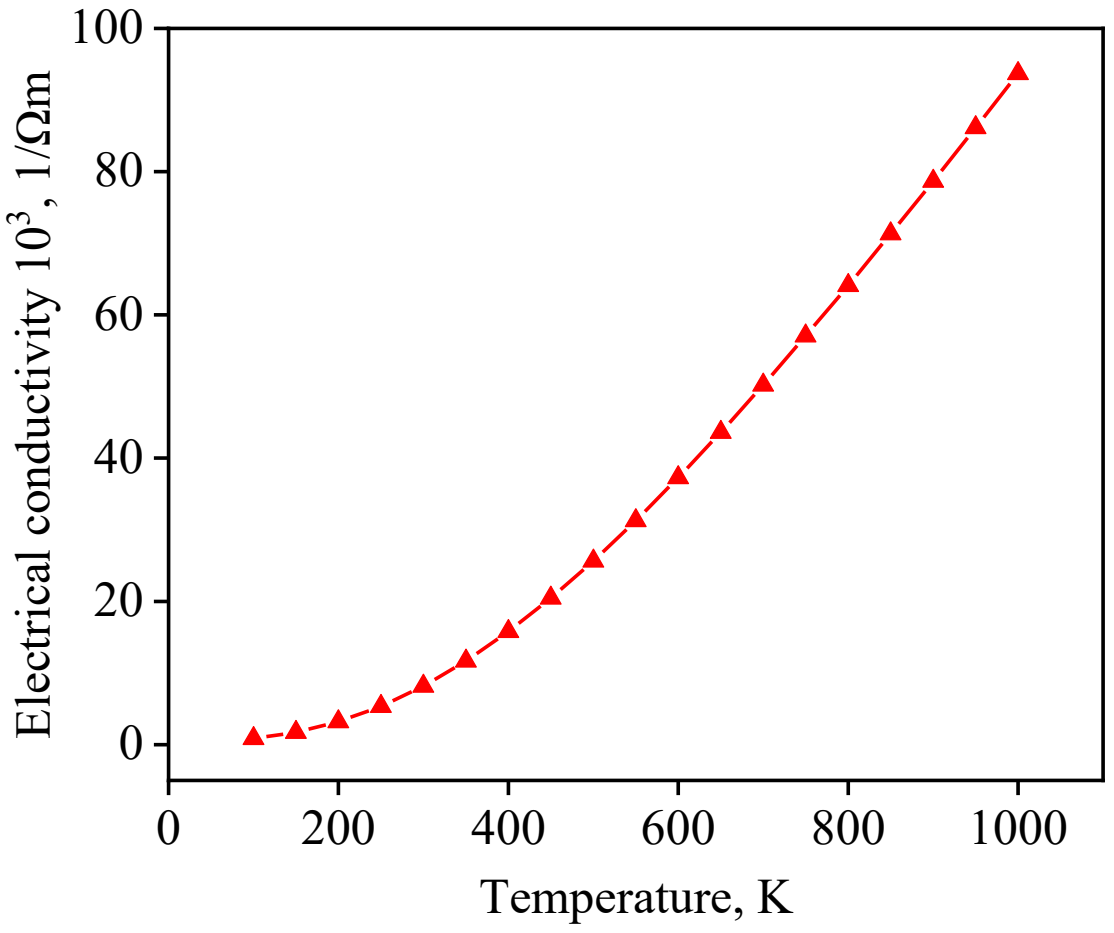

**Fig. 16.** Temperature dependence of electrical conductivity for $Cu_2ZnSnS_4$**.**

The power factor (PF), shown in Figure 17, exhibits a monotonic increase with temperature. As temperature increases, the increase in σ compensates for the decrease in the Seebeck coefficient, leading to an increase in PF. The maximum PF value reaches $5.33 \times 10^{-3}$ W/K$^2$m at 1000 K, indicating the high thermoelectric potential of $Cu_2ZnSnS_4$ in the high-temperature range.

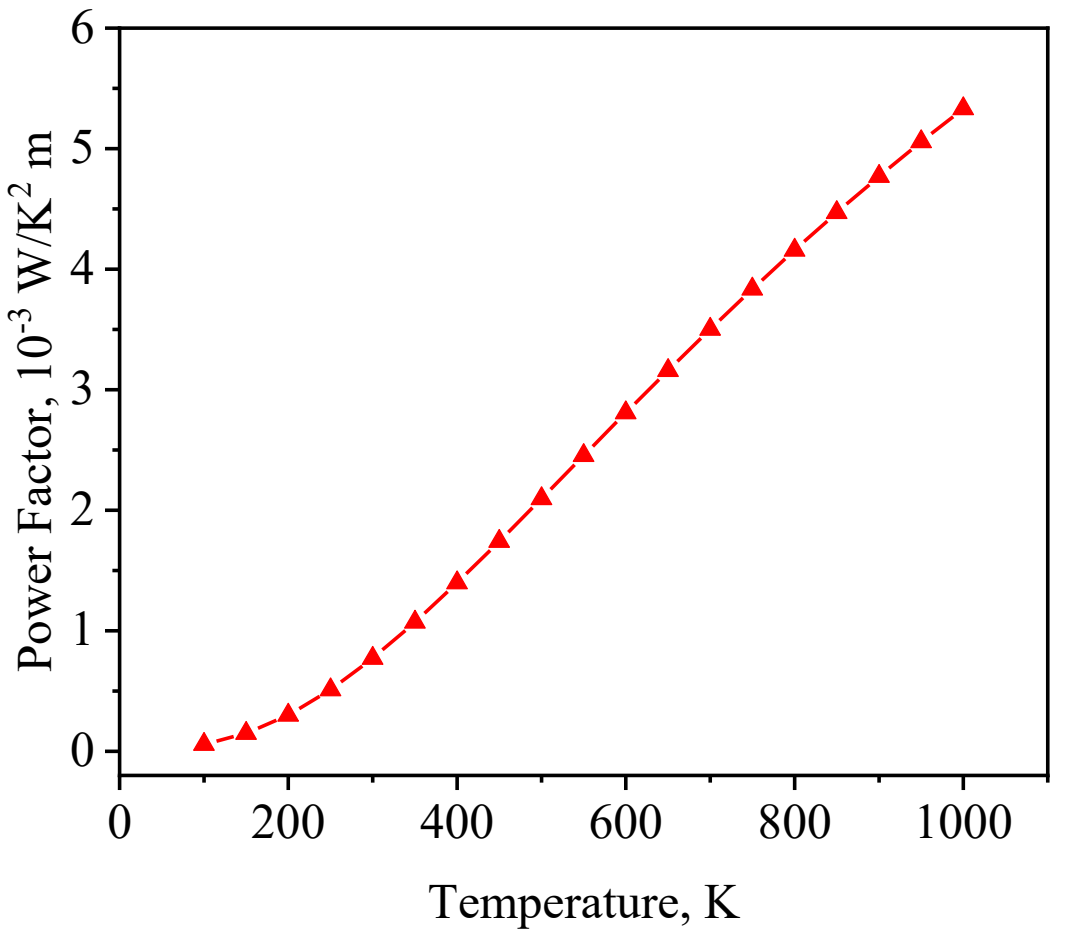

**Fig. 17.** Temperature dependence of the power factor of $Cu_2ZnSnS_4$.

Next, the thermoelectric properties of $Cu_2ZnSnS_4$ - such as the Seebeck coefficient, electrical conductivity, electronic thermal conductivity, and power factor - were comprehensively investigated as a function of chemical potential at temperatures of 300, 600, and 900 K (Figure 20).

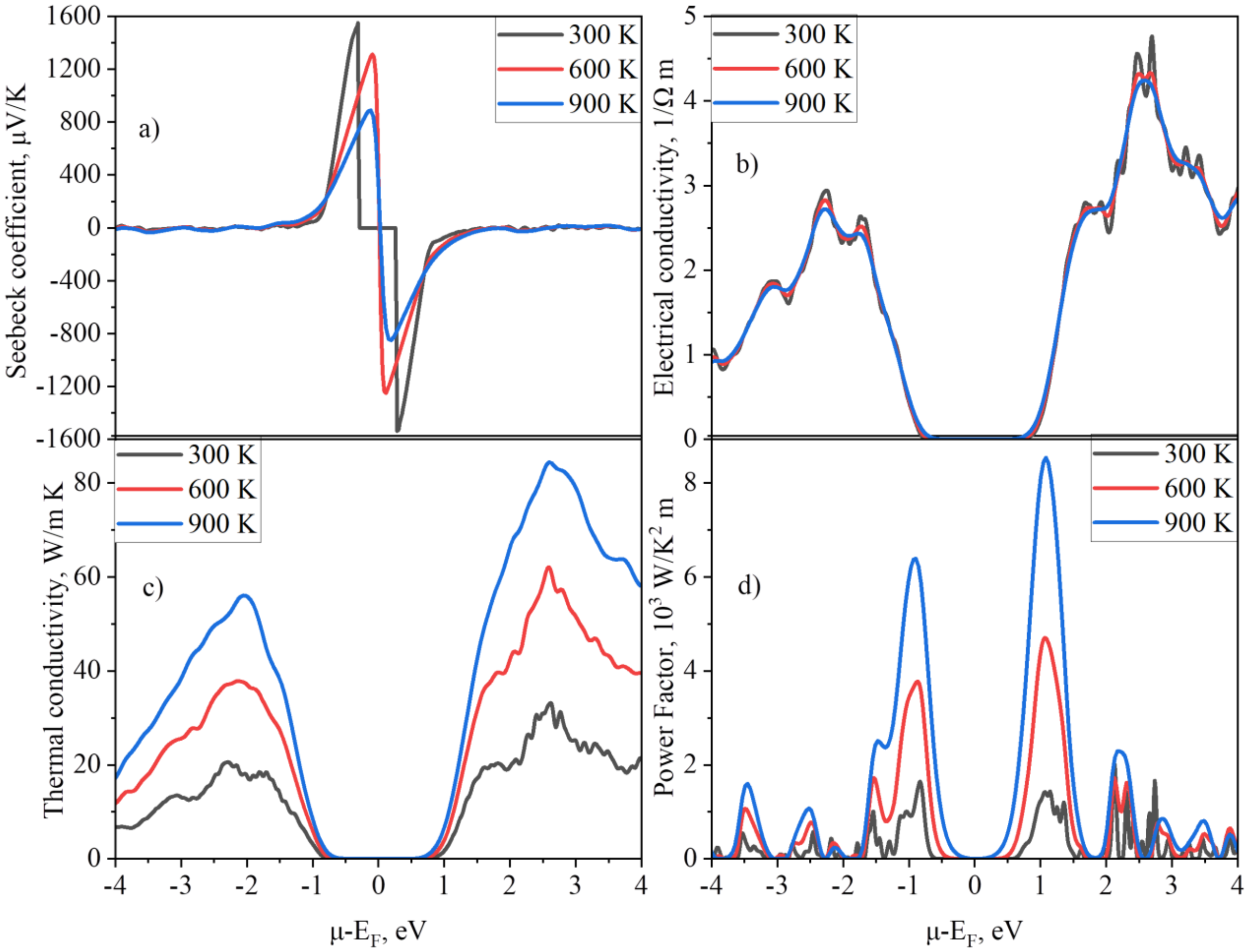

**Fig. 19.** The following parameters were studied as a function of chemical potential: a) Seebeck coefficient, b) electrical conductivity, c) thermal conductivity, and d) power factor.

It can be observed from Figure 19(a) that as the temperature increases from 300 K to 900 K, the peaks decrease in magnitude and shift toward higher values of chemical potential. The maximum Seebeck coefficient occurs at 300 K and is approximately 1550 µV/K. Figure 19(b) illustrates the dependence of electrical conductivity on the chemical potential. It can be seen that in the range of chemical potential corresponding to the band gap, the conductivity is very low. The conductivity increases from zero to an optimal value of 4.75 $\Omega^{-1}m^{-1}$ at an energy level of 2.67 eV. The conductivity is influenced by the transport distribution function, while the effect of temperature is minimal. The electronic thermal conductivity is shown in Figure 19(c). Electron-mediated thermal transport, analogous to electrical conduction, was determined for $Cu_2ZnSnS_4$ in the electron-doped regime. The power factor for kesterite was also calculated and is presented in Figure 19(d). The power factor value is temperature-dependent and increases with rising temperature.

**Conclusion**

The structural, electronic, optical, mechanical, thermodynamic, and thermoelectric properties of kesterite $Cu_2ZnSnS_4$ were systematically investigated using density functional theory. Structure optimization using the SCAN functional was combined with electronic structure calculations using mBJ and HSE06. The results showed the presence of a direct band gap at the G point with a calculated value of 0.88 in mBJ calculations. The band gap value (1.55) obtained using HSE06 is in good agreement with the published experimental value. Hydrostatic compression in the range of 0–30 kbar leads to a gradual contraction of the crystal lattice and an increase in the band gap, demonstrating the sensitivity of the electronic structure to pressure. The calculated elastic properties confirm the mechanical stability of the studied kesterite structure in this range. Moreover, the Pugues coefficient remains above 1.75, indicating plastic behavior under the studied conditions. These results demonstrate that pressure provides the ability to tune the electronic properties while maintaining a mechanically stable structure within the adopted computational model. Phonon spectrum calculations confirm the dynamical stability of $Cu_2ZnSnS_4$. Thermodynamic analysis in the temperature range of 0–1000 K reveals a monotonic decrease in Helmholtz free energy, accompanied by an increase in vibrational entropy. When heated, heat capacity increases rapidly, but at temperatures above 500 K its dependence on temperature weakens. Calculations show that the Seebeck coefficient, after reaching a maximum at 250 K, begins to decrease; however, at 1000 K, it remains approximately 239 $\mu V \cdot K^{-1}$. The increase in electrical conductivity compensates for this decline, leading to an enhancement of the power factor. The dependence of transport coefficients on the chemical potential indicates that the charge-carrier concentration is a critical parameter for further optimization. Future research should focus on charge-carrier scattering processes, doping, and phonon transport in order to derive quantitative performance predictions and evaluate the practical potential of $Cu_2ZnSnS_4$ for thermoelectric energy conversion.

## CRediT authorship contribution statement

**Iskandar Raufzoda:** Writing – original draft, Visualization. **Amondulloi Burhonzoda:** Writing – original draft, Investigation, Methodology. **Dilshod Nematov:** Methodology, Investigation, Conceptualization. **Makhmadzoir Kayumov:** Visualization, Formal analysis, Data curation. **Anushervon Ashurov:** Formal analysis, Data curation. **Saidjafar Murodzoda:** Formal analysis, Data curation. **Sakhidod Sattorzoda:** Formal analysis, Data curation. **Makhsud Islomzoda:** Conceptualization. **Kholmirzo Kholmurodov:** Writing – review & editing, Supervision. **Shuwei Tang:** Writing – review.

## Declaration of competing interest

The authors declare that they have no known competing financial interests or personal relationships that could have appeared to influence the work reported in this paper.

## Funding Acknowledgements

This work was supported by the International Science and Technology Center under Project TJ-0040, and by the Interstate Fund for Humanitarian Cooperation of the CIS Member States through a scientific project funded by the International Nanotechnology Innovation Center of the CIS (Grant No. 22-112).

## Use of AI tools

Artificial intelligence tools were used solely for language polishing, grammatical correction, and minor editorial assistance. All scientific content, data analysis, and interpretation of the results were performed exclusively by the authors. The authors carefully reviewed and verified all AI-assisted text to ensure accuracy, originality, and compliance with ethical publishing standards, including the COPE guidelines.